\documentclass{ijmpi_authors}

\usepackage{epstopdf}

\newcommand{\spion}{NP}
\newcommand{\hdrive}{H_\mathrm{scan}}
\newcommand{\hbdrive}{H_\mathrm{scan}^\mathrm{bi}}

\newcommand{\hmod}{\delta H_\mathrm{m}}

\newcommand{\idrive}{I_\mathrm{scan}}
\newcommand{\bspion}{B_\mathrm{\spion}}

\newcommand{\frf}{f_\mathrm{rf}}
\newcommand{\fl}{f_\mathrm{prec}}

\newcommand{\fmod}{f_\mathrm{m}}
\newcommand{\fspion}{f_\mathrm{\spion}}

\newcommand{\plight}{U_\mathrm{PD}}
\newcommand{\mspion}{M_\mathrm{\spion}}
\newcommand{\sspion}{S_\mathrm{\spion}}
\newcommand{\fdrive}{f_\mathrm{scan}}
\newcommand{\opm}{\mathrm{OPM}}

\begin{document}

\twocolumn[
\title{MPS and ACS with an atomic magnetometer}

\author{Colombo}{Simone}{a,\ast}
\author{Lebedev}{Victor}{a}
\author{Gruji\'c}{Zoran~D.}{a}
\author{Dolgovskiy}{Vladimir}{a}
\author{Weis}{Antoine}{a}

\affiliation{a}{D\'{e}partement de Physique, Universit\'{e} de Fribourg, Chemin du Mus\'{e}e 3, 1700 Fribourg, Switzerland}
\affiliation{\ast}{Corresponding author, email: simone.colombo@unifr.ch}

\maketitle

\begin{abstract}
We show that a single atomic magnetometer in a magnetically unshielded environment can be used to perform magnetic particle spectroscopy (MPS) and AC susceptometry (ACS) on liquid-suspended magnetic nanoparticles.
We demonstrate methods allowing a simultaneous recording of $M(H)$ and d$M$/d$H(H)$ dependences of samples containing down to 1~$\mu$g of iron.
Our results pave the way towards an atomic magnetometer based MPI scanner.
\end{abstract}
]

\section{Introduction}

The technique of Magnetic Particle Imaging (MPI), following its introduction in 2005 \cite{GleichN2005}, has evolved along two major pathways, viz., frequency-space MPI \cite{panagiotopoulos2015} and X-space MPI \cite{GoodwillAM2012}, both approaches having their respective merits and drawbacks.
MPI relies on exciting a magnetic nanoparticle (MNP) sample by a monochromatically oscillating (frequency $\fdrive$) drive field $\hdrive(t)$---referred to as `scan' field in this paper---and detecting the polychromatic time-dependent magnetic induction $\bspion(t){\propto}\mspion$ that arises as a consequence of the nonlinear $\mspion(\hdrive)$ relation.
Most MPI approaches developed so far share the common feature that the detected signal $\sspion(t)$ is recorded by induction coil(s).
Because of Faraday's induction law, one has  $\sspion{\propto}d\bspion(t)/dt{\propto}d\mspion(t)/dt{\propto}n\,\fdrive$, where $n$ denotes the order of the detected overtones, so that signal/noise considerations call for large drive frequencies.
A high-frequency drive favors high speed operation permitting fast volumetric scans of the sample \cite{WeizeneckerAPD2008}.

Frequency-domain operation profits from S/N ratio enhancement by selective bandpass-filtering of the harmonics at known frequencies, but involves demanding calibration procedures.
X-space MPI, on the other hand, allows a simpler interpretation of the recorded (filtered) time series, leading to a computationally less demanding image reconstruction based on the a priori known field free point (or line) position.
Both methods suffer from a direct strong drive field contribution to the detected induction.
%
From the point of view of applying MPI to biological systems, in particular to humans, unwanted neural stimulation and unwanted side effects from excessive SAR (specific absorption rate) become pronounced at elevated frequencies and/or drive coil power \cite{SchmaleIEEEM2015}.

Here we report on the detection of the anharmonic magnetic response of MNPs by a high sensitivity optically pumped atomic magnetometer ($\opm$), and demonstrate that an $\opm$ can be used for Magnetic Particle Spectroscopy (MPS) and AC susceptometry (ACS).
MPS is often referred to as zero-dimensional MPI, and any high-sensitivity MPS method is thus a necessary prerequisite for developing an MPI system.

Laser-driven $\opm$s are compact and versatile instruments, mostly operating at room temperature, that can detect magnetic field changes in the femto- or even sub-femto-Tesla  range.
%
Recent review of various $\opm$ principles and their applications is given in Ref.~\cite{Budker}.
%
%
In the past decade $\opm$s have been deployed for biomagnetism studies, such as magnetocardiography \cite{BisonAPL2009,AlemPMB2015} and magnetoencephalography \cite{SanderBOE2012}.
The suitability of $\opm$s for studying the magnetorelaxation of \textit{blocked MNPs} was demonstrated in recent years \cite{MaserRSI2011,JohnsonJMMM2012,DolgovskiyJMMM2015}.
More recently we have shown (Ref.~\cite{ColomboInprepMH} in this volume) that $\opm$s can also be used for the quantitative measurement of the saturation magnetization, $M_S$, the iron content and particle size distributions of \textit{aqueous MNP suspensions}.
We believe that because of their high sensitivity and large bandwidth (DC up to hundreds of kHz), $\opm$s, when combined with a variant of X-space MPI, have the potential to yield a complementary, low-frequency MPI technique. 

\section{Experimental apparatus}
\begin{figure}[!t]
	\centering
	\includegraphics[width=1\columnwidth]{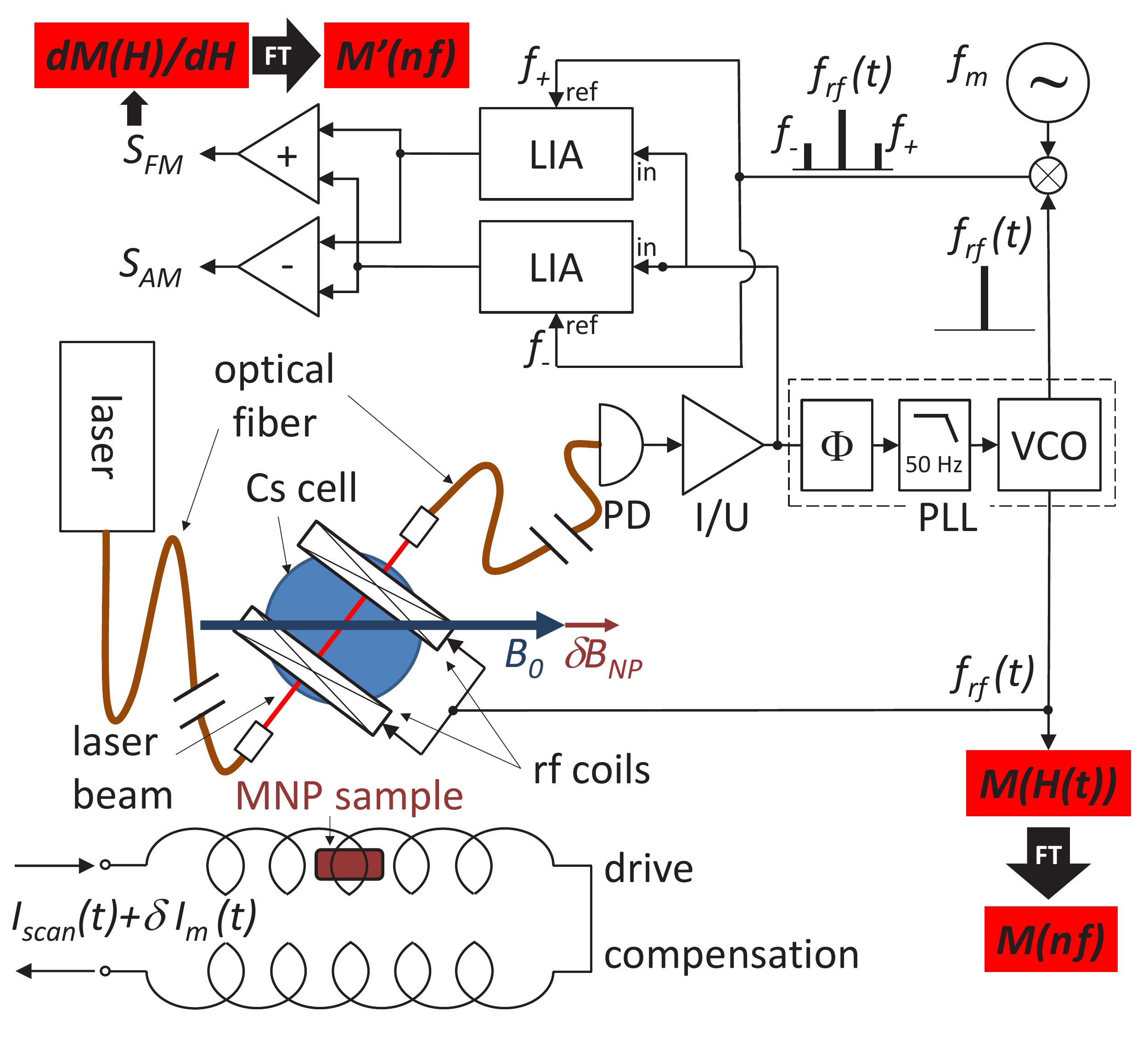}
	\caption{Sketch of the experimental setup.
 LIA: lock-in amplifier; PD: photodiode; I/U: transimpedance amplifier; VCO: voltage-controlled oscillator.
 $f_\pm$ refers to frequencies $f_{rf}\pm f_m$.
 }
	\label{fig:setup1}
\end{figure}

The main components of the apparatus (mounted in a walk-in size double-layer aluminium chamber) are sketched in Figure~\ref{fig:setup1}.
The MNP sample is excited by a periodically oscillating scan field $\hdrive(t)$  produced by a $700$~mm long, $14$~mm diameter solenoid,
next to which an identical, but oppositely poled solenoid reduces residual field at the $\opm$ location (~${\sim}7$~cm above the two solenoids) to $\sim$1~nT per mT of scan field (more details are given in \cite{ColomboInprepMH}).
%
%
%
%

%
The $\opm$ is based on optically detected magnetic resonance in spin-polarized Cs vapor \cite{BisonAPL2009}.
The sensor is operated in a homogeneous bias magnetic field $B_0$ of 27~$\mu$T, produced by large ($\sim$3$\times$3$\times$3~m$^3$) Helmholtz coils, which also cancel the local laboratory field.
We note that the MNP sample is also exposed to that bias field.
The Cs spin polarization in the sensor precesses with a Larmor frequency $\fl$  of 95~kHz in that field.
The precession is driven phase-coherently by a weak (few nT) field oscillating at the `rf' frequency $\frf$ generated by a low bandwidth ($\sim$50~Hz) phase-locked loop (PLL, Fig.~\ref{fig:setup1}) ensuring that the magnetic resonance condition $\frf{= }\fl$ is maintained when $\fl$, i.e., $|\vec{B}_0|$ varies.
The driven Cs spin precession impresses a modulation (at $\frf$) on the detected light power.
%
%
The $B_0$ field information is thus encoded in terms of the frequency $\frf{=}\gamma_F|\vec{B}_0|$, where $\gamma_F{\approx}$~3.5~Hz/nT for Cs.
%

%
%
%
%
%
%

%
\section{Measurements and results}\label{sec:results}
Small MNP-induced field changes $|\delta\vec{B}_\mathrm{\spion}(t)|{\ll}|\vec{B}_0|$ in the total field at the sensor position $\vec{B}_\mathrm{tot}(t)$ yield corresponding $\opm$ frequency changes
\begin{eqnarray}
\delta\fspion(t)=f(t){-}f_0=&\gamma_F\left(|\vec{B}_\mathrm{tot}(t)|-|\vec{B}_0|\right)\nonumber\\
=&\gamma_F\delta\vec{B}_\mathrm{\spion}(t)\cdot\hat{B}_0\,.
\label{eq:gyromag}
\end{eqnarray}
To first order in $\delta\bspion$ the $\opm$ signal is thus determined by the projection $\delta\bspion(t){\equiv}\vec{B}_\mathrm{\spion}(t)\cdot\hat{B}_0$ of the field of interest onto the bias field, making the $\opm$ an effective vector component magnetometer.
The sensitivity to MNP signals will be maximized by having $\vec{B}_0\parallel\delta\vec{B}_\mathrm{\spion}$.

The HF2LI lock-in amplifier delivers a voltage $S_{\delta f}(t)=\alpha\,\delta\fspion(t)$, which allows retrieving the corresponding detected induction change (Fig.~\ref{fig:setup1}).

%
%
\subsection{Magnetization curve $M(H)$}\label{ssec:MH}

\begin{figure}[h]
	\centering
	\includegraphics[width=1\columnwidth]{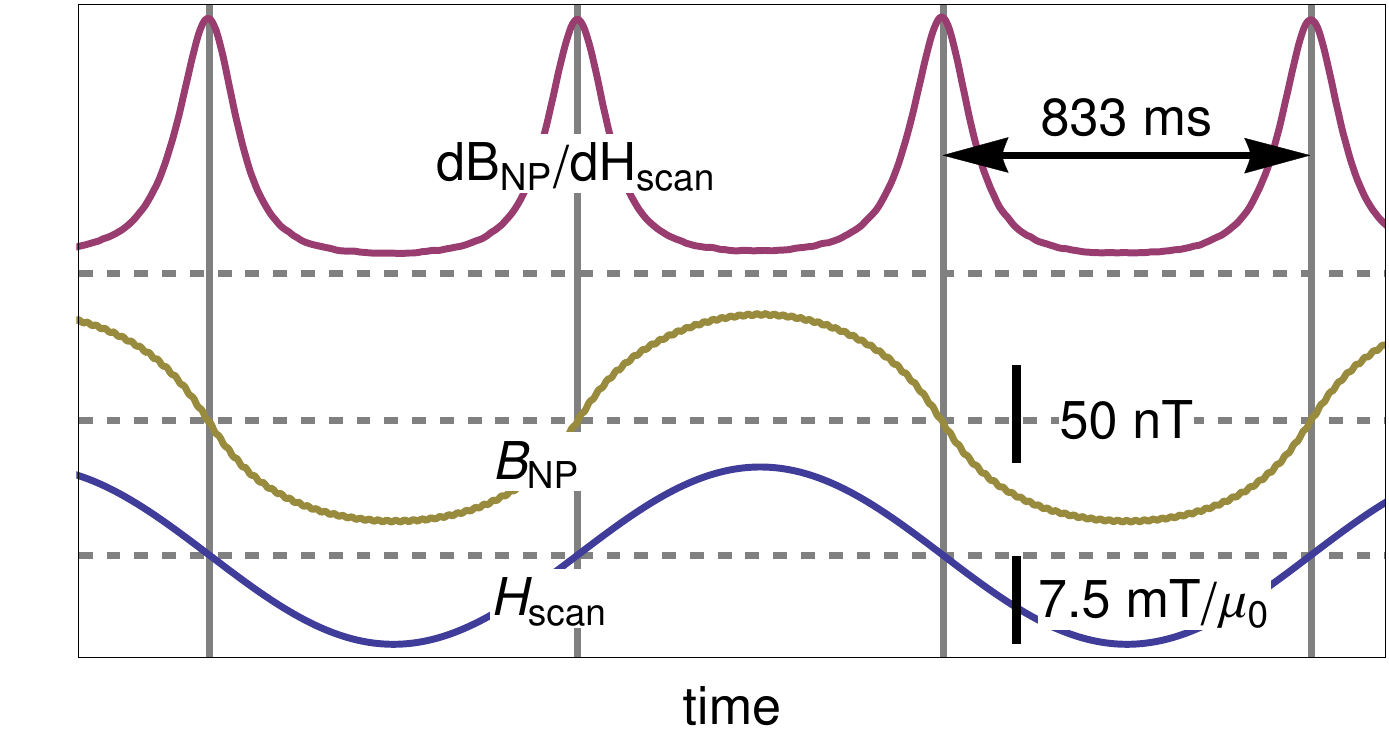}
	\caption{From bottom to top: Time series of the scan field $\hdrive(t)$, the corresponding induced  $\delta\bspion(t){\propto}\mspion(t)$, and d$\bspion/$d$\hdrive(t){\propto}$d$\mspion/$d$\hdrive(t)$ dependencies.
The scan frequency was 600~mHz and the data shown are unfiltered raw data containing 533 data points per period.}
	\label{fig:timeseries}
\end{figure}

Similarly to the method described in \cite{ColomboInprepMH} we excite the sample by a field $\hdrive(t)$ of amplitude $\sim 15$~mT$_{pp}$/$\mu_0$ that sinusoidally oscillates at a frequency $\fdrive$ of 600~mHz.
We record time series (sampled at a rate of 320~S/s) of $\idrive(t)$ and the induced signals $\delta\bspion(t){=}S_{\delta f}(t)/(\alpha\,\gamma_F)$.
These time-space results are shown as the lower two traces of Fig.~\ref{fig:timeseries}.
The recording was done on a 500~$\mu$l EMG${-}707$ sample containing 3.4~mg of iron (more details on the sample given in \cite{ColomboInprepMH}).

\begin{figure}[t]
	\centering
	\includegraphics[width=1\columnwidth]{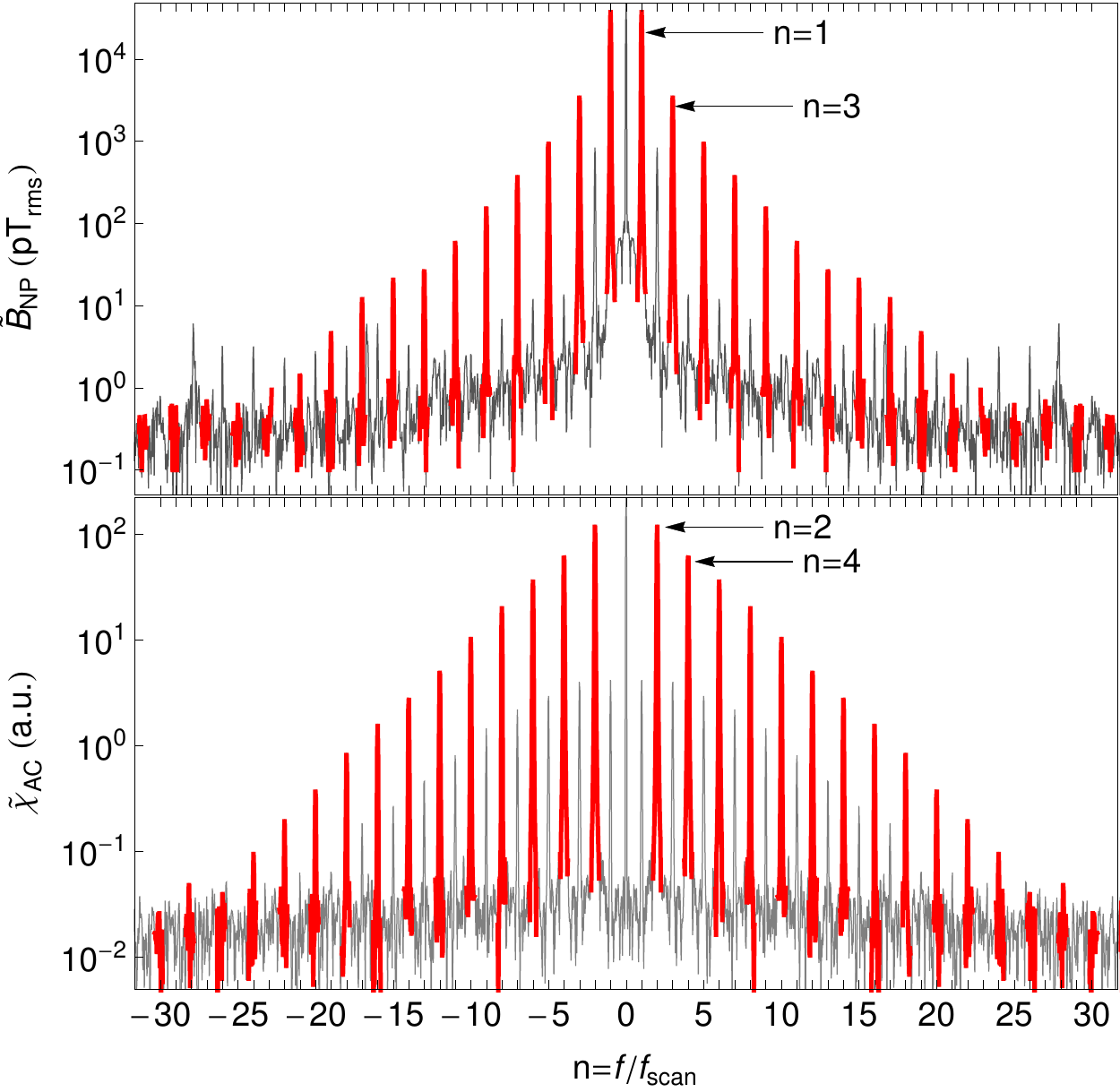}
	\caption{
Fourier transforms of 30 cycles ($\sim$50~s recording time) of $\bspion(t)$ and d$M$/d$H(t)$ data as shown in Fig.~\ref{fig:timeseries}. Top: Magnetic particle spectrum (MPS). Bottom: Fourier spectrum of AC susceptibility.
}
	\label{fig:dMdH}
\end{figure}

%
%
%
%
%
Performing the Fourier transform of data from 30 consecutive scan cycles yields the harmonics spectrum
\begin{equation}
\widetilde{B}_\mathrm{NP}(f/\fdrive)=\mathcal{F}\left[\delta\bspion(t)\right]=\mathcal{F}\left[\frac{S_{\delta f}(t)}{\alpha\,\gamma_F}\right]
\end{equation}
that can be rescaled to magnetization units by
\begin{equation}
\mspion=\frac{4\pi R^3}{\mu_0 V_s}\,\widetilde{B}_\mathrm{NP}\,,
\label{eq:BtoM}
\end{equation}
 where $V_s$ is the sample volume and $R$ the sample-magnetometer spacing (7~cm).%
The top graph of Fig.~\ref{fig:dMdH} shows such a  Fourier spectrum.
When rescaled to unity bandwidth, the noise floor in the figure represent a  power spectral density of $\sim$4~pT/$\sqrt{\mathrm{Hz}}$ which limits the  detectable number of harmonics in the $M(H)$ signals to $\approx23$ for 3.4~mg of iron.
The noise pedestal underlying the low-frequency Fourier components reflects low-frequency noise and drifts of the magnetic field at the sensor location.
We draw attention to the fact that, conversely to conventional MPS methods, our technique gives also access to the fundamental frequency of the MNPs' magnetic response.
In relation to the latter statement we also note that here we derive the harmonics spectrum of $\widetilde{M}_\mathrm{NP}(f/\fdrive)$ from a direct measurement of $\mspion\left(H(t)\right)$, while conventional MPS (and  MPI) devices do the opposite, i.e., reconstruct $\mspion(H)$ from recorded $\widetilde{M}_\mathrm{NP}(f/\fdrive)$ values.
%
%
%

\subsection{AC susceptibility $\chi_{AC}${=}d$M$/d$H$}\label{ssec:dMdH}
We have extended the method discussed above for measuring $M(H)$ curves towards the direct and simultaneous recording of the derivative, i.e.,  d$M$/d$H(H)$-dependence.
For this we excite the sample with a bichromatic superposition of fields
\begin{equation}
\hbdrive(t)=\hdrive\cos\left(2\pi \fdrive t\right)+\hmod\cos\left(2\pi \fmod t\right)\,,
\end{equation}
with a scan field of amplitude $\hdrive\sim$15~mT$_{pp}/\mu_0$ and a weaker modulation field $\hmod{<}$2~mT$_{pp}/\mu_0$.
The quasi-static scan at frequency $\fdrive{=}$1~Hz${\ll}\fmod{=}$753~Hz implies that at any given time $t_0$, the MNPs' magnetization is given by  $\mspion(t_0){=}\sum_{n=0} m_n(t_0) \cos(n~2\pi\fmod t)$, where $m_0\equiv M_{\spion}\left[\hdrive(t_0)\right]$, and where
%
%
%
%
\begin{eqnarray}
m_{n>0}(t_0){=}4\int\limits_{\fmod t_0}^{\fmod t_0+\frac{1}{2}}M_{\spion}\left[\hbdrive(t)\right]\cos(n 2\pi\fmod t)\mathrm{d}(\fmod t)\nonumber\\
{=}\frac{2\hmod}{n\pi}\int\limits_{-1}^{1}M'_{\spion}\left[\hbdrive(t_0;x)\right]\sqrt{1{-}x^2}U_{n-1}(x)\mathrm{d}x\,.
\end{eqnarray}
%
%
%
In the last expression $U_{j}$ is the $j{-}$th Chebyshev polynomial of the second kind.
For $\hmod{<}H_k$, $H_k$ being the sample's saturation field, the harmonic amplitudes $m_{n{>}1}$ are negligible, and the first harmonic reduces to
%
%
\begin{equation}
m_1(t_0)=\hmod\,\frac{d\mspion\left[\hdrive(t_0)\right]}{dH}
 {\equiv}\hmod~M'_{\spion}\left[\hdrive(t_0)\right]\,\label{eq:m_1}\,.
\end{equation}
At each time $t_0$, the magnetization component oscillating at $\fmod$ is thus proportional to the derivative of the MNPs' $M(H)$ dependence (Langevin function).

We extract a signal proportional to $m_1\left(\hdrive(t_0)\right)$ in the following way:
%
%
The sample's magnetization component $m_0$ (varying at the slow frequency $\fdrive$) and the sample's magnetization component $m_1$ (oscillating at $\fmod$) produce magnetic induction fields $\bspion(t_0)$ and $\delta \bspion(t_0)cos(2\pi\fmod t)$, respectively, that add to the offset field $B_0$ at the sensor location.

The magnetometer signal is a photocurrent oscillating at a frequency proportional to the modulus of the total induction field
\begin{equation}
B_\mathrm{tot}(t){=}B_0{+}\bspion(t_0){+}\delta\bspion\cos(2\pi\fmod t)\,,
\end{equation}
where $|\bspion|{\ll}|B_0|$.
One sees that the problem of inferring the amplitude of the $m_1(t)$ component is a problem of FM spectroscopy.
The corresponding time-dependent photodiode signal is given by
\begin{equation}
\plight(t)=A(t)\cos\left[2 \pi \frf\,t + \frac{f_\mathrm{p}}{\fmod}\sin\left(2\pi \fmod t\right)\right]\,,
\label{eq:FM_1}
\end{equation}
where $f_p{=}\gamma_F\hmod$.
Since the PLL is tracking slow variations $\bspion$ of $B_0$ field, we have omitted, in Eq.~\ref{eq:FM_1}, the contribution $\gamma_F \bspion(\fdrive t_0)$ to $\frf{=}\gamma_F B_0$.
We extract the sideband amplitudes by the sideband demodulation technique illustrated in Fig.~\ref{fig:setup1}, which thus gives simultaneous access to both $dM/dH(H)$ and $M(H)$.

The top trace of Fig.~\ref{fig:timeseries} shows the derivative signal d$\bspion/$d$\hdrive(t)$, which---after calibration by Eq.~\ref{eq:BtoM}---is equivalent to $d\mspion/dH$.
The Fourier spectrum of 30 cycles of d$\bspion/$d$\hdrive(t)$ data is shown as lower graph in Fig.~\ref{fig:dMdH}.
Comparison of the two Fourier spectra reveals the superior power of FM-spectroscopy:
While the direct $M(H)$ method is sensitive to drift and low frequency noise of the `background' field $B_0$ at the sensor (as evidenced by the noise pedestal under the upper spectrum in Fig.~\ref{fig:dMdH}), the derivative spectrum is insensitive to low-frequency changes of the carrier frequency $\frf$.

While the $\widetilde{B}_\mathrm{NP}(f)$ spectrum is dominated by odd Fourier components, even frequency  components dominate the
\begin{equation}
\widetilde{\chi}_{AC}(f)\propto\mathcal{F}\left[\frac{\mathrm{d}\bspion(t)}{\mathrm{d}H}\right]
\end{equation}
spectrum.

The Fourier spectra show some artifacts.
The upper graph of Fig.~\ref{fig:dMdH} contains a series of even harmonics arising from field components $\delta \vec{B}_{\perp}(t)$ perpendicular to $\vec{B}_0$.
Since $|\vec{B}_\mathrm{tot}(t)|=\sqrt{(B_{0}+\delta B_{\mathrm{NP}}(t))^2+\delta B_{\perp}^2(t)}$ the signal contains terms oscillating at even harmonics, the dominant one being at 2$f_{\mathrm{scan}}$.
In the lower graph of Fig.~\ref{fig:dMdH} the odd harmonics are mainly due to the fact that $\hdrive$ does not oscillate around zero, but rather around the average value $\langle\hdrive\rangle{=}B_0/\mu_0$ of 27~$\mu$T/$\mu_0$ (bias field).

In a series of dilution experiments we have demonstrated the proportional scaling of the $M(H)$ \cite{ColomboInprepMH} and d$M$/d$H$ signals with iron content.
%
%
Based on the lower graph of Fig.~\ref{fig:dMdH} that was recorded with 3.4~mg of iron, we estimate that our current detection limit in a recording time of 50~s is $\sim$700~ng of iron.

\section{Conclusions and outlook}\label{sec:summary}
We have demonstrated that an atomic magnetometer in an unshielded environment can be used for a direct quantitative measurement of MPS and ACS spectra of magnetic nanoparticles in the \mbox{sub-kHz} frequency range.
At current stage the method allows absolute iron content determinations at a sub-$\mu$g level.
The low-frequency scans give access to the response of large particles that are hydrodynamically blocked at the often used 25~kHz modulation frequency.
%
%

Because of the relatively slow ($\approx$1~Hz) scan speed, the $M(H)$-recording is perturbed by magnetic field instabilities in our unshielded environment and suffers from low-frequency noise of the deployed scan current supply.
As demonstrated in Ref.~\cite{BisonAPL2009}, a first- or second-order $\opm$ gradiometer arrangement is able to improve the  magnetometric sensitivity by up to two orders of magnitude.
Work towards this goal is in progress.

We have also demonstrated a superior method that allows AC susceptometry (d$M$/d$H$) recordings based on an FM operation mode of the magnetometer.
%
%
The latter approach gives direct access to the first harmonic response of the MNPs, information that is missing in standard MPI methods.

We note that the derivative curve d$M$/d$H$, when multiplied by a linear gradient field d$H$/d$x$, yields the point-spread function d$M$/d$x$, which---together with signal/noise considerations---defines the spatial resolution for X-space MPI~\cite{GoodwillAM2012}.
Based on the excellent signal quality demonstrated above we now pursue the goal of designing a 2D X-space MPI scanner based on induction field detection by atomic magnetometers.
In contrast to conventional pick-up coils that do not `feel' inhomogeneous DC fields, the atomic sensor performance rapidly degrades in gradient fields, so that the main challenge will be the design of a gradient field generator that will be compatible with a high sensitivity mode of operation of the magnetometer.
%
%
\vspace{11mm}

\textbf{Acknowledgements}.
This work was supported by Grant No. $200021\_149542$ from the Swiss National Science Foundation.

\bibliography{AM-MPS}

\end{document}